\documentclass[times,5p,twocolumn,final]{elsarticle}
\usepackage{amsmath}
\usepackage[hidelinks]{hyperref}
\usepackage{subcaption}
\usepackage{booktabs}
\usepackage{tikz}
\usepackage{wrapfig}
\usepackage[inline]{enumitem}
\usepackage{siunitx}
\usepackage{multirow}
\usepackage[ruled,linesnumbered]{algorithm2e}

\tikzset{font=\scriptsize}
\usetikzlibrary{arrows.meta}

\DeclareMathOperator*{\argmin}{arg\,min}

\begin{document}

\begin{frontmatter}

\title{Diverse Part Synthesis for 3D Shape Creation}

\author{Yanran Guan\corref{cor1}}
\cortext[cor1]{Corresponding author}
\ead{yanran.guan@carleton.ca}
\author{Oliver van Kaick}
\ead{oliver.vankaick@carleton.ca}
\address{School of Computer Science, Carleton University, Canada}

\begin{abstract} 
Methods that use neural networks for synthesizing 3D shapes in the form of a part-based representation have been introduced over the last few years. These methods represent shapes as a graph or hierarchy of parts and enable a variety of applications such as shape sampling and reconstruction. However, current methods do not allow easily regenerating individual shape parts according to user preferences. In this paper, we investigate techniques that allow the user to generate multiple, diverse suggestions for individual parts. Specifically, we experiment with multimodal deep generative models that allow sampling diverse suggestions for shape parts and focus on models which have not been considered in previous work on shape synthesis. To provide a comparative study of these techniques, we introduce a method for synthesizing 3D shapes in a part-based representation and evaluate all the part suggestion techniques within this synthesis method. In our method, which is inspired by previous work, shapes are represented as a set of parts in the form of implicit functions which are then positioned in space to form the final shape. Synthesis in this representation is enabled by a neural network architecture based on an implicit decoder and a spatial transformer. We compare the various multimodal generative models by evaluating their performance in generating part suggestions. Our contribution is to show with qualitative and quantitative evaluations which of the new techniques for multimodal part generation perform the best and that a synthesis method based on the top-performing techniques allows the user to more finely control the parts that are generated in the 3D shapes while maintaining high shape fidelity when reconstructing shapes.

\end{abstract}


\end{frontmatter}

\section{Introduction}\label{sec:intro}

Synthesizing 3D shapes is an important research problem in computer graphics, which has applications in the creation of content for games, simulations, and virtual worlds. The latest methods for automatic or semi-automatic synthesis of 3D shapes make use of deep neural networks, generating shapes either through the sampling of latent or noise vectors~\cite{chen2019learning,li2021sp}, diffusion models~\cite{zeng2022lion,hui2022neural}, or through language guidance~\cite{chen2018text2shape,liu2022towards,sanghi2022clip}. Many of these methods synthesize shapes as a whole in the form of point clouds~\cite{li2021sp,zeng2022lion}, volumes~\cite{sanghi2022clip}, or implicit functions~\cite{chen2019learning,hui2022neural,li2022learning}.

On the other hand, shapes can also be represented as a composition of parts, either in the form of a graph or hierarchy that explicitly encodes the shape's structure~\cite{mitra2013structure}. This representation has advantages in modeling and analysis, such as the possibility of modifying and exchanging individual parts or annotating shapes with semantic information. A recent line of work has used part-based representations for 3D shape synthesis with neural networks, proposing methods that can be employed for a variety of useful tasks such as shape synthesis from latent representations~\cite{schor2019componet,mo2019structurenet,wu2019sagnet,wu2020pq,niu2022rim}, part-based shape reconstruction~\cite{dubrovina2019composite,petrov2023anise}, and shape interpolation~\cite{yang2022dsg,wu2020pq,mo2019structurenet}.

\begin{figure}[t]
    \centering
    \includegraphics{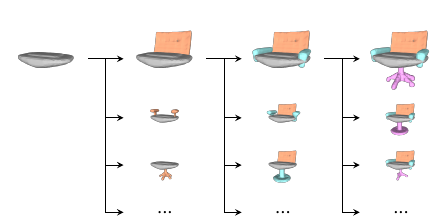}
    \caption{Our method for incremental shape synthesis based on diverse part suggestions. Starting from an initial part (in gray), our method iteratively synthesizes new part suggestions and connects them to the existing part(s) to compose a coherent shape. Various, distinct part suggestions are proposed at each iteration (each column), and the user selects one part to proceed to the next iteration (top row). Note that the geometry of the parts is also synthesized by the method.}
    \label{fig:teaser}
\end{figure}

Despite the various uses of these methods, one limitation of the current work is that it is difficult to control the generation of individual shape parts. That is, if the user would like to modify a specific part that has been generated, the existing methods do not provide a specific mechanism for generating multiple meaningful suggestions to replace the part. Such a functionality is ideal in an iterative modeling framework where the user starts with an initial shape suggested by the modeling system and refines its details to progressively approximate a mental design. In the current methods, shapes are either represented as latent vectors that encode entire shapes~\cite{wu2020pq,niu2022rim} or do not offer mechanisms for regenerating suitable latent codes for individual parts while ensuring the variety of the parts and their consistency with the current shape~\cite{schor2019componet,dubrovina2019composite,petrov2023anise}. A few works have introduced methods for disentangling parts from structure~\cite{yang2022dsg}, which allow interpolating geometry or structure independently. However, control and resampling of individual parts is not explicitly enabled by the disentangled representation.

In this paper, we investigate the development of a method that synthesizes shapes in a part-based representation while allowing the user to generate a variety of distinct suggestions for individual parts. Specifically, we perform a comparative study of multimodal deep generative models that enable to synthesize multiple, diverse part suggestions to compose a shape. The use of a multimodal generative model is especially important to ensure the diversity of the suggested parts. Deep generative models are known for facing problems such as mode collapse, which results in the networks operating practically in a deterministic manner, where highly similar outputs are generated for a same input. Therefore, in this work, we experiment with various multimodal generative models and conduct a comparative study of these models, identifying the most proficient ones in generating suggested parts with both shape fidelity and diversity. Specifically, we evaluate mixture density networks (MDNs)~\cite{bishop1994mixture}, as well as conditional generative models implemented with generative adversarial networks (GANs)~\cite{goodfellow2014generative}, generator networks trained with implicit maximum likelihood estimation (IMLE)~\cite{li2018implicit}, and denoising diffusion probabilistic models (DDPMs)~\cite{ho2020denoising}.

To compare all the models within the same synthesis framework, we introduce an effective method for synthesizing 3D shapes in a part-based representation and evaluate all the part suggestion techniques with this same synthesis method. In our synthesis framework, the user constructs a shape in an incremental manner (Figure~\ref{fig:teaser}). First, the user selects an initial part or partial assembly of parts for building a shape, and the method suggests suitable new parts that can be added to the partial assembly. The user can then select a part, and the method automatically connects the part to the partial assembly. This process continues in an iterative manner until the shape reaches a desired number of parts. Note that this incremental construction approach can also be used for automatic sampling of shapes with diverse structures by automating the part selection step with randomized choices.

The incremental shape synthesis is enabled by a method composed of three main neural networks:
\begin{enumerate*}[label=(\roman*)]
\item a decoder that generates shapes in an implicit representation based on a latent vector,
\item a spatial transformer~\cite{jaderberg2015spatial} that positions the synthesized parts in their target locations, and
\item a multimodal generative model that suggests latent vectors of suitable parts that can be connected to a given partial assembly.
\end{enumerate*}
In addition, an autoencoder (AE) architecture ensures that a latent space of shape parts is learned by the method.

We show with qualitative and quantitative evaluations, compared to other 3D generative baselines, that our overall method for part-based shape synthesis is able to generate shapes in a part representation where the synthesized shapes have good reconstruction accuracy and reasonable novelty according to common evaluation metrics.

We summarize our contributions as follows:
\begin{itemize}
\item We perform a comparative study of multimodal generative models for suggesting diverse parts for shape synthesis.
\item We demonstrate that the conditional models implemented with IMLE and DDPMs provide the best overall part suggestions in terms of visual quality and diversity.
\item We introduce the first part-based synthesis method that explicitly allows regenerating diverse part suggestions with a thoroughly evaluated deep learning-based approach.
\end{itemize}

\section{Related work}\label{sec:related}

In this section, we review the literature most related to our work.

\subsection{Part-based shape synthesis}\label{related:synthesis}

Shape synthesis by data-driven part assembly has been an area of active research in computer graphics for a few decades, given the flexibility of modeling shapes as a composition of parts~\cite{funkhouser2004modeling,mitra2013structure}. Earlier part-based assembly methods make use of a variety of techniques such as part suggestion via probabilistic reasoning~\cite{chaudhuri2011probabilistic,kalogerakis2012probabilistic,sung2017complementme} or set evolution~\cite{xu2012fit,guan2022fame}. These methods typically start from a predefined set of parts and combine the parts to generate new shapes. Recent methods for part-based synthesis make use of deep neural networks, encoding shapes and their parts in a variety of representations, such as point clouds~\cite{schor2019componet}, voxel sets~\cite{dubrovina2019composite,li2020learning}, and implicit fields~\cite{wu2020pq,petrov2023anise}. Part-based assembly methods based on neural networks can be roughly classified into four groups.

The first group of methods assume that a set of parts is given as input and mainly provide recombinations of these parts for generating new shapes, similarly as the earlier part-based assembly methods. For example, in their ComplementMe system, Sung et al.~\cite{sung2017complementme} pose the problem of suggesting parts to be added to an incomplete partial assembly as a retrieval problem solved with a neural network. Similarly, Sung et al.~\cite{sung2018learning} encode the interchangeability or complementarity of shape parts in a dual embedding, while Huang et al.~\cite{huang2020generative} introduce a method for reassembling parts through dynamic graph learning. To improve the visual fidelity of the assembled parts, Yin et al.~\cite{yin2020coalesce} reconstruct the part joints as implicit surfaces. Although these methods can be used for generating shapes with new combinations of parts, the geometry of the parts is limited to the existing collection, and thus parts with varied geometry cannot be generated.

The second group of methods synthesizes the part structure and geometry of the parts independently, thus enabling the generation of novel shape structures and their parts. This is typically achieved with the combination of a part generation network and a part placement network, which also inspired the architecture of our method. For example, Dubrovina et al.~\cite{dubrovina2019composite} introduce a neural network for part-based reconstruction based on a part decoder and localization network. Schor et al.~\cite{schor2019componet} propose a method composed of a part decoder network and a composition network, generating shapes in a part-based representation by randomly sampling part and composition latent codes, while Li et al.~\cite{li2020learning} also use latent codes for generating parts as volumes but predict part placement from the synthesized volumes. ANISE~\cite{petrov2023anise} reconstructs shapes with a part-based reconstruction pipeline based on part and position codes and an implicit representation of parts. Although the methods in this group allow generating or reconstructing shapes in a part-based presentation, they provide limited or no functionality for easily replacing or regenerating specific shape parts. All parts are either assumed to be independent and fully exchangeable~\cite{schor2019componet}, or the methods do not provide mechanisms for regenerating specific parts codes from partial shapes, given their main goal of shape reconstruction~\cite{dubrovina2019composite,petrov2023anise}.

Moreover, the third group of methods synthesize the part structure and individual parts of a shape based on a single latent code. The method either generates the entire shape from the vector or iteratively generates a sequence of vectors from the initial one to guide the generation of parts. This is achieved with a global-to-local GAN alongside a part refiner by Wang et al.~\cite{wang2018global}, a hierarchical graph representation in StructureNet~\cite{mo2019structurenet}, a joint latent space of structure and geometry in SAGNet~\cite{wu2019sagnet}, sequential generation of a part structure in PQ-NET~\cite{wu2020pq}, or a sequence of latent spaces encoded from a shape's hierarchy in LSD-StructureNet~\cite{roberts2021lsd}. These methods are quite versatile given that a complex shape can be generated from a single latent code. However, since individual parts are not associated to specific entries of the latent codes, the regeneration of specific parts dependent on other parts of the shape cannot be easily achieved.

Finally, the fourth group of methods propose to learn part-based representations in an unsupervised or self-supervised manner. RIM-Net~\cite{niu2022rim} learns recursive implicit fields for encoding shapes in a hierarchical structure. DSG-Net~\cite{yang2022dsg} encodes shapes in a representation that disentangles geometry and structure, such that the structure and geometry of shapes can be sampled or interpolated independently. SPAGHETTI~\cite{hertz2022spaghetti} supports direct editing of 3D models by learning a representation of shape parts and their transformations. The advantage of these methods is that labeled data is not required for training, since the loss ensures that a part decomposition is obtained in an unsupervised manner. However, the part segmentation discovered by the methods may not be ideal for certain shapes, and the part decomposition is not guaranteed to fit the user's intended semantic abstraction of the shapes.

\subsection{Deep generative models with multimodality}\label{related:multimodal}

A common problem in machine learning is how to synthesize results with diversity, given that common generative models operate in practice in a monomodal manner, synthesizing nearly identical outputs for a given input. For example, research in image translation has reported this effect in conditional image generation~\cite{isola2017image}. In the case of GANs, the monomodal generation arises due to the problem of mode collapse, where the training process maps a certain group of inputs to a single output that should be predicted.

Research in this area has attempted to address the diversity problem in different manners. One group of methods predict a distribution of latent vectors, rather than a single vector, for a given input. For example, the distribution can be encoded in the form of a mixture of Gaussians~\cite{bishop1994mixture}, so that multiple, distinct vectors can be sampled from the distribution. This type of approach has been used for shape generation~\cite{sung2017complementme,jahan2021semantics}. Since the distribution is learned independently from the embedding, it is not guaranteed to be contained in the latent manifold, which implies that decoding samples of the distribution may lead to suboptimal results.

Another line of work introduces conditional generative models for multimodal generation, e.g., conditioning the input to a GAN~\cite{mirza2014conditional} or to a variational autoencoder (VAE)~\cite{sohn2015learning}. Wu et al.~\cite{wu2020multimodal} combine a VAE with a GAN for multimodal generation in shape completion. However, despite the multimodal output, the quality of the results suffers from the VAE encoding and decoding. To prevent mode collapse incurred by GANs, IMLE has been proposed by Li and Malik~\cite{li2018implicit}. The main idea of this method is to match each training sample to a generated sample during training, so that the network takes into consideration all possible modes in the training set. This helps to alleviate the problem of mode collapse when learning generative models. Recently, DDPMs proposed by Ho et al.~\cite{ho2020denoising} have emerged as a novel powerful deep generative model. DDPMs can be made conditional by using the condition information as an additional input through the diffusion pathway~\cite{nichol2021improved}.

The sampling procedure of the aforementioned conditional models is based the idea of an implicit generative model~\cite{mohamed2016learning}, i.e., adding a random noise vector $\mathbf{n}\sim\mathcal{N}(0,\mathbf{I})$ to a deterministic parameterized transformation $F_\theta$, so that, given the conditional input $\mathbf{c}$, the predicted sample $\hat{\mathbf{x}}$ is drawn from $F_\theta$ as:
\begin{equation}\label{conditional-sampling}
    \hat{\mathbf{x}}=F_\theta(\mathbf{c},\mathbf{n}).
\end{equation}

Conditional models have been applied to tasks such as super-resolution~\cite{chen2018fsrnet,li2020multimodal,rombach2022high,saharia2023image} and image synthesis from a given layout~\cite{isola2017image,li2019diverse,li2020multimodal,rombach2022high,wang2022semantic} and also to shape completion~\cite{dai2017shape,wu2020multimodal,zhou20213d,arora2022multimodal,cheng2023sdfusion,li2023diffusion,chu2024diffcomplete} and shape generation from text~\cite{chen2018text2shape,liu2022towards,zeng2022lion,cheng2023sdfusion,li2023diffusion}. Note that these approaches predict or complete entire shapes at a time.

In the next section, we provide more details on how we use these multimodal generative models in our method. We also present a comparison of these models in Section~\ref{sec:results}, to show which models perform the best in providing multiple, distinct, and high-quality suggestions for a given partial shape assembly.

\section{Shape synthesis with part suggestions}\label{sec:method}

\begin{figure*}
    \centering
    \includegraphics{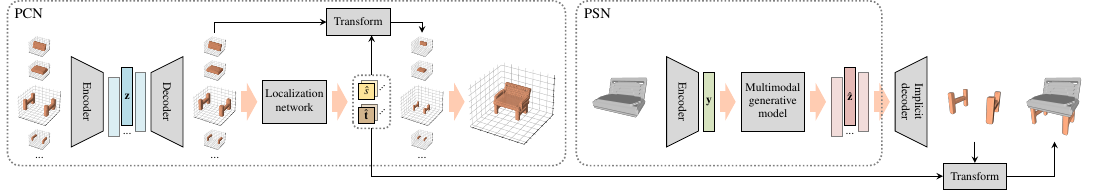}
    \caption{Our method for 3D shape synthesis with part suggestions comprises two main deep neural networks. During the training phase, the PCN receives a set of normalized and transformed shape parts. The PCN first encodes the normalized shape parts into a latent vector $\mathbf{z}$ and then learns the affine transformation parameters that transform the normalized parts to their target positions in order to compose a coherent shape. The PSN receives a partial assembly and learns to generate the latent vector $\hat{\mathbf{z}}$ that represents the shape part complementing the partial assembly. After the networks are trained, given a partial assembly, we use the PSN to generate a set of latent samples and pass them to an implicit decoder to produce the suggested parts. Then, we use the PCN to predict the affine transformations to connect the suggested parts to the partial assembly.}
    \label{fig:overview}
\end{figure*}

In this section, we introduce our method for synthesizing 3D shapes in a part-based representation, which we use for evaluating the multimodal part suggestion techniques. Our method is composed of two main deep neural networks:
\begin{enumerate*}[label=(\roman*)]
\item a part composition network (PCN) that learns a latent space of shape parts and the affine transformation applied to each part and
\item a part suggestion network (PSN) that learns to generate the latent codes of the possible parts given an incomplete partial assembly.
\end{enumerate*}
The PSN suggests parts with one of the multimodal generative models that we cover in our study. Moreover, to achieve a higher visual quality of the synthesized shapes, we use an implicit decoder~\cite{chen2019learning}, which decodes latent vectors generated by the PSN into implicit fields representing shape parts. We illustrate in Figure~\ref{fig:overview} the architecture of the PCN and PSN and the process of part suggestion and placement.

\begin{algorithm}[t]
    \caption{Iterative synthesis}\label{alg:iterative_synthesis}
    \DontPrintSemicolon
    \KwIn{$\mathbf{P}$, $K$}
    \KwOut{$\mathcal{A}$}
    \SetKwProg{Fn}{function}{}{end}\SetKwFunction{FIterativeSynthesis}{IterativeSynthesis}
    \Fn{\FIterativeSynthesis{\textup{\texttt{$\mathbf{P}$,$K$}}}}{
        $\mathcal{A}\gets\{\mathbf{P}\}$\;
        $i\gets1$\;
        \While{$i\leq K$}{
            \ForEach{$\mathbf{A}\in\mathcal{A}$}{
                $\mathcal{A}^\prime\gets\emptyset$\;
                $\mathbf{y}\gets\texttt{Encoder($\mathbf{A}$)}$\;
                $\hat{\mathcal{Z}}\gets\texttt{MultimodalGenerativeModel($\mathbf{y}$)}$\;
                \ForEach{$\hat{\mathbf{z}}\in\hat{\mathcal{Z}}$}{
                    $\hat{\mathbf{P}}\gets\texttt{Decoder($\hat{\mathbf{z}}$)}$\;
                    $\hat{s},\hat{\mathbf{t}}\gets\texttt{LocalizationNetwork($\hat{\mathbf{P}}$)}$\;
                    $\hat{\mathbf{P}}\gets\texttt{ImplicitDecoder($\hat{\mathbf{z}}$)}$\;
                    $\hat{\mathbf{P}}^\prime\gets\texttt{Transform($\hat{\mathbf{P}}$,$\hat{s}$,$\hat{\mathbf{t}}$)}$\;
                    $\mathbf{A}^\prime\gets\mathbf{A}+\hat{\mathbf{P}}^\prime$\;
                    $\mathcal{A}^\prime\gets\mathcal{A}^\prime\cup\{\mathbf{A}^\prime\}$\;
                }
            }
            $\mathcal{A}\gets\texttt{UserSelection($\mathcal{A}^\prime$)}$\;
            $i\gets i+1$\;
        }
        \Return{$\mathcal{A}$}\;
    }
\end{algorithm}

We first train the PCN to derive the latent space of shape parts and then use the latent codes of shape parts to train the PSN for generating multimodal part suggestions. The details of training the PCN and PSN are discussed in the following subsections. Once all the network modules of the PCN and PSN are trained, new shapes can be synthesized iteratively with incrementally suggested shape parts. Given an initial shape part or partial assembly as input, the PSN generates possible latent codes representing the shape parts that can complement the current partial assembly. Then, given a part suggestion, we transform the decoded shape part into its target position based on the prediction of affine transformations learned by the PCN. Thus, to generate a complete shape, we start from an initial part and iteratively add more parts to the assembly based on the suggestions from the PSN and decoding and placement of the PCN. This process can be carried out in a user-in-the-loop manner, where a user selects one or more of the suggested parts at each iteration, or in a randomized manner, where one or multiple suggestions are automatically selected to compose a set of randomly sampled shapes. We detail this iterative shape synthesis process via incremental part suggestion in Algorithm~\ref{alg:iterative_synthesis}. The synthesis is carried out by the function \texttt{IterativeSynthesis($\mathbf{P}$,$K$)}, which takes an initial shape part $\mathbf{P}$ and a predefined number of iterations $K$ as input and outputs a set $\mathcal{A}$ containing the final synthesized shapes (assemblies). In Algorithm~\ref{alg:iterative_synthesis}, each of the functions, i.e., \texttt{Encoder}, \texttt{Decoder}, \texttt{LocalizationNetwork}, \texttt{Transform}, and \texttt{ImplicitDecoder}, corresponds to a network module presented in Figure~\ref{fig:overview} (in gray), and \texttt{UserSelection} represents the process of users' visual inspection and selection of one or multiple plausible partial assemblies at the end of each iteration.

\subsection{Part composition network (PCN)}\label{method:pcn}

The PCN takes a voxelized shape part as input and outputs the predicted affine transformation parameters of the part, i.e., the scale and translation, and the transformed shape part. The PCN consists of two training units: a volumetric AE that learns a latent representation for each shape part and a spatial transformer network (STN)~\cite{jaderberg2015spatial} that first predicts the affine transformation parameters of parts and then scales and translates the reconstructed parts to their target positions to compose a coherent shape. Note that we do not include rotations in the transformation since the training data is aligned consistently.

\paragraph{Training set} We denote the training set of the PCN with $N$ samples as $\mathcal{T}_\text{PCN}=\{(\mathbf{P}_i,\mathbf{P}^\prime_i,s_i,\mathbf{t}_i)\}^N_{i=1}$, where each training sample contains a normalized and a transformed shape part $\mathbf{P}_i$ and $\mathbf{P}^\prime_i$, both axis-aligned and voxelized with a resolution of $64\times64\times64$, and a scalar $s_i$ along with a 3D vector $\mathbf{t}_i$, respectively describing the uniform scale and translation of the affine transformation that transforms $\mathbf{P}_i$ to $\mathbf{P}^\prime_i$.

\paragraph{Network architecture} We feed the input shape part, represented as a set of $64\times64\times64$ voxels, into the AE, which comprises an encoder and a decoder symmetric to each other. The encoder downsamples the input to $4\times4\times4$ using $5$ convolution layers followed by a batch normalization and generates the latent vector $\mathbf{z}$ using a fully connected (FC) layer activated with a sigmoid function. In our implementation, we use a latent space of $128$ dimensions. The decoder transfers $\mathbf{z}$ back into a shape part represented as a volume using the inverse architecture of the encoder. 

The STN has two modules: a localization network and a transformation module. The reconstructed shape part produced by the AE is passed to the localization network, which regresses the transformation parameters (a scale and a translation) using $4$ convolution layers followed by a batch normalization and $4$ FC layers. The transformation module utilizes the learned transformation parameters to generate an affine grid which is then applied to the reconstructed shape part through bilinear sampling, warping the transformed shape part to its correct position in the full shape.

\paragraph{Loss function} The training objective of the PCN can be decomposed into minimizing two losses, similar to the scheme adopted by Hu et al.~\cite{hu2018predictive}:
\begin{enumerate*}[label=(\roman*)]
\item the reconstruction loss of the AE, i.e., the mean squared error (MSE) between the input and reconstructed shape parts $\mathbf{P}_i$ and $\hat{\mathbf{P}}_i$, and
\item the STN loss, i.e., the MSE between the ground truth (GT) and predicted transformed shape parts $\mathbf{P}^\prime_i$ and $\hat{\mathbf{P}}^\prime_i$, added to the difference between the GT transformation parameters ($s_i$ and $\mathbf{t}_i$) and learned parameters ($\hat{s}_i$ and $\hat{\mathbf{t}}_i$).
\end{enumerate*}
Therefore, the loss function of the PCN is a linear combination of the aforementioned losses, which for each training sample is written as:
\begin{equation}\label{pcn-loss}
\begin{split}
    \mathcal{L}_\text{PCN}&=\mathcal{L}_\text{AE}+\mathcal{L}_\text{STN}\\
    &=\operatorname{MSE}(\mathbf{P}_i,\hat{\mathbf{P}}_i)+\operatorname{MSE}(\mathbf{P}^\prime_i,\hat{\mathbf{P}}^\prime_i)+d(s_i,\hat{s}_i)+d(\mathbf{t}_i,\hat{\mathbf{t}}_i),
\end{split}
\end{equation}
where $d(\cdot,\cdot)$ is a distance metric. We use the Euclidean distance (ED) in our experiments.

\subsection{Part suggestion network (PSN)}\label{method:psn}

Given a partial assembly comprising one or more shape parts, the PSN predicts multiple latent vectors, where each vector corresponds to a suggested shape part that can be connected to the current partial assembly. Our PSN consists of two components: the encoder network pretrained in the PCN that encodes the partial assemblies into latent codes and a deep generative model that enables multimodal part suggestions. We experiment with various deep generative models with multimodality in the implementation of the PSN, including the MDN, conditional GAN (cGAN), conditional IMLE (cIMLE), and conditional DDPM (cDDPM).

\paragraph{Training set} We denote the training set of the PSN with $M$ samples as $\mathcal{T}_\text{PSN}=\{(\mathbf{A}_i,\mathbf{z}_i)\}^M_{i=1}$. Each training sample comprises a partial assembly $\mathbf{A}_i$ and a latent vector $\mathbf{z}_i$ of the shape part, learned by the PCN, that complements the partial assembly. To generate a partial assembly, we randomly sample one or more transformed parts from a shape and combine them into one assembly. Then, for each partial assembly, we randomly select a normalized part from the same shape that complements the partial assembly and feed the selected shape part to the PCN to derive the part's latent vector. In addition, to simplify the training process, we feed each partial assembly $\mathbf{A}_i$ to the encoder pretrained in the PCN, to derive a $128$-dimensional latent code $\mathbf{y}_i$ that is used as the input of the multimodal generative model.

We discuss the different multimodal generative models in the next subsections.

\subsubsection{Mixture density network (MDN)}\label{method:mdn}

An MDN combines a neural network with a mixture density model. It learns a regression that maps the input data to a mixture of multiple Gaussian distributions, modeling the conditional probability distribution of the target data.

\paragraph{Network architecture} Each input vector $\mathbf{y}_i$ is processed with a feed-forward neural network composed of $3$ FC layers. The output of the network is a mixture model of parts' latent vectors conditioned on the input vectors. The mixture model contains $h$ Gaussian distributions (we use $h=4$) and is represented as $3$ vectors: one $h$-dimensional vector storing the mixing coefficient for each distribution and two vectors with dimension $h\times128$ storing the mean and standard deviation that represent each Gaussian distribution in the mixture model.

\paragraph{Loss function} The training objective of the regression is to produce, for each partial assembly code $\mathbf{y}_i$, a mixture model that maximizes the sampling probability of $\mathbf{z}_i$. The mixture model is composed of $h$ Gaussian distributions and its probability density is represented by a linear combination of Gaussian kernel functions:
\begin{equation}\label{mdn}
    P(\mathbf{z}_i\mid\mathbf{y}_i)=\sum^h_{j=1}w_j\phi_j(\mathbf{z}_i\mid\mathbf{y}_i),
\end{equation}
where $h$ is the number of distributions in the mixture, $w_j$ represents the mixing coefficient of the $j$th kernel, and $\phi_j(\mathbf{z}_i\mid\mathbf{y}_i)$ is the probability density function (PDF) of the target $\mathbf{z}_i$ for the $j$th kernel. Thus, the loss function of the network is defined as minimizing the negative logarithm of the PDFs of the kernels, i.e.,
\begin{equation}\label{mdn-loss}
    \mathcal{L}_\text{MDN}=-\log P(\mathbf{z}_i\mid\mathbf{y}_i).
\end{equation}

\subsubsection{Conditional GAN (cGAN)}\label{method:cgan}

A GAN consists of two components: a generator network and a discriminator network. Both components are trained simultaneously in a competitive process. The generator learns to generate fake but plausible samples from random noise vectors, while the discriminator learns to distinguish the generator's fake outputs from the real ones, through penalizing the generator for producing implausible results.

\paragraph{Network architecture} Both the generator and discriminator consist of $3$ FC layers. The generator takes a conditional input by concatenating the partial assembly code $\mathbf{y}_i$ to a random noise drawn from $\mathcal{N}(0,\mathbf{I})$ and generates a fake part code. Similarly, we concatenate the condition $\mathbf{y}_i$ to a part latent code as the input to the discriminator, which outputs a logit value describing the probability that the part code is a real one.

\paragraph{Loss function} We denote the generator and the discriminator as $G(\cdot,\cdot)$ and $D(\cdot,\cdot)$ and the randomly sampled noise vectors as $\mathbf{n}_i$. The loss function of the cGAN is written as:
\begin{equation}\label{cgan-loss}
    \mathcal{L}_\text{cGAN}=\log D(\mathbf{y}_i,\mathbf{z}_i)+\log\left(1-D(\mathbf{y}_i,G(\mathbf{y}_i,\mathbf{n}_i))\right).
\end{equation}

\subsubsection{Conditional IMLE (cIMLE)}\label{method:cimle}

\begin{figure}
    \centering
    \begin{subfigure}{0.495\linewidth}
        \centering
        \includegraphics{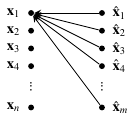}
        \caption{Mode collapse}
        \label{fig:imle-mode-collapse}
    \end{subfigure}
    \hfill
    \begin{subfigure}{0.495\linewidth}
        \centering
        \includegraphics{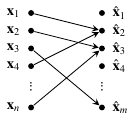}
        \caption{IMLE}
        \label{fig:imle-imle}
    \end{subfigure}
    \caption{Illustration of how IMLE prevents mode collapse: (a)~The common training scheme for generative models maps each training sample $\hat{\mathbf{x}}_j$ to the closest real sample $\mathbf{x}_i$, which can lead to some modes ($\mathbf{x}_i$) not being represented by any sample; (b)~IMLE prevents mode collapse by ensuring a mapping from each real sample $\mathbf{x}_i$ to at least one generated sample $\hat{\mathbf{x}}_j$.}
    \label{fig:imle}
\end{figure}

IMLE can be used to train generative models according to the following formulation. We denote a set of $n$ data samples as $\{\mathbf{x}_i\}^n_{i=1}$ and the parameters of a probability distribution $\mathcal{D}$ as $\theta$. Given $m$ i.i.d. random samples $\{\hat{\mathbf{x}}_j\}^m_{j=1}$ drawn from $\mathcal{D}$, where $m\geq n$, IMLE optimizes the following objective function to find the optimal parameters $\hat{\theta}$:
\begin{equation}\label{imle}
    \hat{\theta}=\argmin_\theta\mathbb{E}_{\{\hat{\mathbf{x}}_j\}^m_{j=1}}\left[\sum^n_{i=1}\min^m_{j=1}d(\mathbf{x}_i,\hat{\mathbf{x}}_j)\right],
\end{equation}
where $d(\cdot,\cdot)$ refers to a distance metric. Note that the sum on the right side of the formula considers the minimum for each sample $\mathbf{x}_i$, ensuring that each training sample is matched to a sample drawn from the distribution, which helps to prevent mode collapse, as illustrated in Figure~\ref{fig:imle}.

\paragraph{Network architecture} For each partial assembly code $\mathbf{y}_i$, we randomly sample $h$ noise vectors drawn from $\mathcal{N}(0,\mathbf{I})$. Note that $h$ is a hyper-parameter and we use $h=4$. Then, we concatenate each individual random noise vector to $\mathbf{y}_i$, to derive the conditioned input. Finally, we feed the concatenated vectors to a generator, which uses $5$ FC layers to generate $h$ fake latent samples.

\paragraph{Loss function} We denote the generator as $G(\cdot,\cdot)$ and the randomly sampled noise vectors as $\mathbf{n}_{i,j}$ ($j\in\{1,\dots,h\}$). The partial assembly code $\mathbf{y}_i$, which is the conditional input, along with a noise vector $\mathbf{n}_{i,j}$, are fed to the generator to generate a fake sample, i.e.,
\begin{equation}\label{cimle-fake}
    \hat{\mathbf{z}}_{i,j}=G(\mathbf{y}_i,\mathbf{n}_{i,j}).
\end{equation}
The network is trained to minimize the MSE between the GT latent vector $\mathbf{z}_i$ and its nearest generated sample $\hat{\mathbf{z}}_{i,k}$ according to the $k$th noise $\mathbf{n}_{i,k}$:
\begin{equation}\label{cimle-loss}
    \mathcal{L}_\text{cIMLE}=\operatorname{MSE}(\mathbf{z}_i,\hat{\mathbf{z}}_{i,k}),
\end{equation}
with
\begin{equation}\label{cimle-k}
    k=\argmin_{j\in\{1,\dots,h\}}d(\mathbf{z}_i,\hat{\mathbf{z}}_{i,j}),
\end{equation}
so that $\mathcal{L}_\text{cIMLE}$ captures the IMLE objective. We use ED as the distance metric $d(\cdot,\cdot)$.

\subsubsection{Conditional DDPM (cDDPM)}\label{method:cddpm}

The diffusion process, which produces latent codes $\mathbf{x}^1$ through $\mathbf{x}^T$ within $T$ steps, is defined as gradually and iteratively adding Gaussian noise to the original data $\mathbf{x}^0$, i.e.,
\begin{equation}\label{q-sample-1}
    q(\mathbf{x}^1,\dots,\mathbf{x}^T\mid\mathbf{x}^0)=\prod^T_{t=1}q(\mathbf{x}^t\mid\mathbf{x}^{t-1}),
\end{equation}
where
\begin{equation}\label{q-sample-2}
    q(\mathbf{x}^t\mid\mathbf{x}^{t-1})=\mathcal{N}(\mathbf{x}^t;\sqrt{1-\beta_t}\mathbf{x}^{t-1},\beta_t\mathbf{I}),
\end{equation}
with $\beta_t\in(0,1)$ being the variance of the Gaussian noise added in time step $t$ ($t\in\{1,\dots,T\}$). DDPMs approximate the reverse diffusion process at each step using a neural network parameterized by weights $\theta$, i.e.,
\begin{equation}\label{p-sample}
    p_\theta(\mathbf{x}^{t-1}\mid\mathbf{x}^t)=\mathcal{N}(\mathbf{x}^{t-1};\mu_\theta(\mathbf{x}^t,t),\Sigma_\theta(\mathbf{x}^t,t)).
\end{equation}

\paragraph{Network architecture} Given a part latent code $\mathbf{z}_i$ and its corresponding noise $\mathbf{z}^t_i$ obtained after applying the diffusion process $t$ times, we use a U-Net~\cite{ronneberger2015unet} as the backbone model to learn the mapping from $\mathbf{z}^t_i$ to $\mathbf{z}^{t-1}_i$ at each time step $t$. The U-Net consists of a contraction path as an encoder and an expansion path as a decoder. In our U-Net structure, the contraction path has $4$ blocks of convolutional layers that performs downsampling. The expansion path, symmetrically, uses $4$ layer blocks performing upsampling. The U-Net is time-conditional by taking an additional input $t$. In our implementation, the U-Net is also conditioned on the partial assembly code $\mathbf{y}_i$, by concatenating $\mathbf{y}_i$ to $\mathbf{z}^t_i$.

\paragraph{Loss function} We denote the U-Net model as $\epsilon_\theta(\cdot,\cdot,\cdot)$, based on the reparameterization~\cite{ho2020denoising}:
\begin{equation}\label{ddpm-reparameterization}
    \mathbf{x}^t(\mathbf{x}^0,\epsilon)=\sqrt{\bar{\alpha}_t}\mathbf{x}^0+\sqrt{1-\bar{\alpha}_t}\epsilon,
\end{equation}
with
\begin{equation}\label{ddpm-alpha}
    \bar{\alpha}_t=\prod^t_{r=1}(1-\beta_r).
\end{equation}
We adopt the simplified training objective of DDPMs as introduced by Ho et al.~\cite{ho2020denoising} in our cDDPM:
\begin{equation}\label{cddpm-loss}
    \mathcal{L}_\text{cDDPM}=\operatorname{MSE}(\epsilon,\epsilon_\theta(t,\mathbf{y}_i,\mathbf{z}^t_i)),
\end{equation}
where $\epsilon\sim\mathcal{N}(0,\mathbf{I})$ and $t$ is uniformly sampled from $\{1,\dots,T\}$.

\begin{figure*}
    \centering
    \begin{subfigure}{0.06\linewidth}
        \centering
        \includegraphics{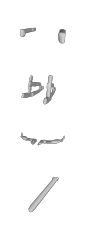}
        \caption*{}
        \label{fig:comparison-initial}
    \end{subfigure}
    \quad
    \hfill
    \begin{subfigure}{0.22\linewidth}
        \centering
        \includegraphics{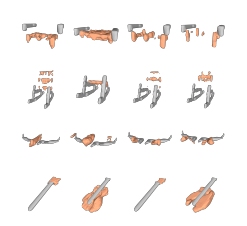}
        \caption{MDN}
        \label{fig:comparison-mdn}
    \end{subfigure}
    \hfill
    \begin{subfigure}{0.22\linewidth}
        \centering
        \includegraphics{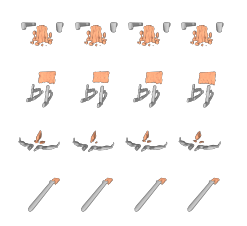}
        \caption{cGAN}
        \label{fig:comparison-cgan}
    \end{subfigure}
    \hfill
    \begin{subfigure}{0.22\linewidth}
        \centering
        \includegraphics{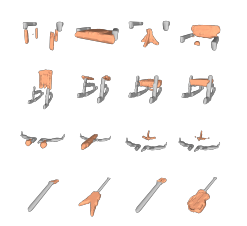}
        \caption{cIMLE}
        \label{fig:comparison-cimle}
    \end{subfigure}
    \hfill
    \begin{subfigure}{0.22\linewidth}
        \centering
        \includegraphics{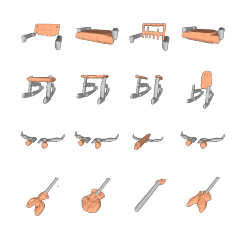}
        \caption{cDDPM}
        \label{fig:comparison-cddpm}
    \end{subfigure}
    \caption{Comparison of the parts suggested by MDN, cGAN, cIMLE, and cDDPM, given the initial parts in gray on the left.}
    \label{fig:comparison}
\end{figure*}

\section{Experimental setup}\label{sec:experiment}

In this section, we introduce our experimental methodology. Implementation details of our experiments can be found in our code\footnote{\url{https://github.com/IsaacGuan/Diverse-Part-Synthesis}}.

\paragraph{Experimental data} In our experiments, we use $500$ chairs, $500$ airplanes, and $500$ guitars from the ShapeNet dataset with semantic segmentation~\cite{chang2015shapenet,yi2016scalable}. We also experiment with training our networks with synthetic shapes generated through a procedural modeling library (PML)~\cite{ali2021evaluation}. We create a set of $500$ chairs using the PML. All the chairs and airplanes in our experiments are composed of $3$ or $4$ semantic parts, and all the guitars are composed of $3$ semantic parts. To better evaluate the quality of the synthesized shapes, we partition each category into training and test sets according to a $4:1$ split. Note that, in general, our method is suitable for shapes that can be decomposed into semantic parts, such as man-made shapes.

\paragraph{Optimizer and hyper-parameters} All the neural networks employed in our experiments are trained using the Adam optimizer~\cite{kingma2015adam}. For the PCN, we experimentally set the learning rate of the AE to $10^{-4}$ and the learning rate of the STN to $10^{-6}$. We employ a two-stage training scenario for the PCN, where the AE and STN are first updated simultaneously for $1000$ epochs, and then we train the STN for another $500$ epochs. We employ different training settings for different versions of PSNs. We train the MDN for $1000$ epochs using a learning rate of $10^{-4}$, the cGAN for $2000$ epochs using a learning rate of $10^{-5}$, the cIMLE for $500$ epochs using a learning rate of $10^{-4}$, and the cDDPM for $5\times10^5$ epochs using a learning rate of $8\times10^{-5}$.

\paragraph{Machine configuration and timing} We trained the neural networks using an NVIDIA GeForce RTX 2080 Ti GPU with $\SI{11}{\giga\byte}$ of memory and CUDA version 11.3. For each iteration of part suggestion, the time required to sample latent codes varies among different models. The MDN takes on average $\SI{0.73}{\second}$ to predict the Gaussian distributions. The cGAN, cIMLE, and cDDPM take on average $\SI{83}{\milli\second}$, $\SI{69}{\milli\second}$, and $\SI{25}{\second}$, respectively, to generate the latent codes. To generate a mesh from each latent code, the inference time of the implicit decoder is around $\SI{2.3}{\second}$, predicting the affine transformation parameters by the PCN takes $\SI{0.46}{\second}$, applying the affine transformation to an implicit field takes $\SI{0.52}{\second}$, and mesh creation with marching cubes~\cite{lorensen1987marching} takes $\SI{11}{\milli\second}$.

\section{Results}\label{sec:results}

In this section, we show qualitative and quantitative results of our comparative study based on different implementations of the PSN. Moreover, we present various shape synthesis results using our proposed pipeline.

\subsection{Comparison of multimodal generative models}\label{res:compare}

We first present a comparative study of the different implementations of the PSN.

Figure~\ref{fig:comparison} gives representative examples of the parts suggested by the different methods for the same initial part. The initial parts are, from top to bottom, chair arms from the PML chairs, chair legs from the ShapeNet chairs, airplane wings from the ShapeNet airplanes, and a guitar neck from the ShapeNet guitars. We observe that the parts sampled from the MDN are blurry and noisy, and no diversity is observed in the cGAN-generated parts (possibly due to mode collapse). However, both cIMLE and cDDPM yield suggested parts with good visual quality and variety, with cIMLE providing slightly better semantic diversity in the generated parts. In the examples of chair arms and airplane wings, the part suggestions generated by cIMLE exhibit variations across all possible semantic labels of chair and airplane parts, while cDDPM is only able to generate chair seats and backs and airplane bodies and engines, respectively.

Table~\ref{tab:distance-comparison} numerically evaluates the diversity in these examples. We compute and compare the distances between all the pairs of parts suggested by each method for each initial part. We use ED, the chamfer distance (CD), and the earth mover's distance (EMD) as the distance metrics, and we present the average of the pairwise distances of the three metrics for each method. All the distances are computed based on parts generated as binary voxel grids with resolution $64\times64\times64$. For computing CD and EMD, we convert the generated parts in volumetric format into point clouds. Specifically, we first sample a point for each occupied voxel and then rescale the sampled points into a unit sphere. We see from Table~\ref{tab:distance-comparison} that the parts generated by cIMLE possess the most diversity, since larger pairwise distances indicate that the shapes are more distinct from each other.

\begin{table}
\caption{Numerical evaluation of the diversity in the suggested parts generated by different methods, where CD and EMD are multiplied by $10^2$.}
\label{tab:distance-comparison}
\small
\centering
\begin{tabular}{ lccc }
\toprule
Method & ED & CD & EMD \\
\midrule
MDN & $27.19$ & $33.87$ & $22.01$ \\
cGAN & $7.546$ & $6.762$ & $4.823$ \\
cIMLE & $\mathbf{40.20}$ & $\mathbf{47.33}$ & $\mathbf{27.83}$ \\
cDDPM & $38.12$ & $40.90$ & $22.53$ \\
\bottomrule
\end{tabular}
\end{table}

\begin{table}
\caption{Quantitative evaluation of the diversity and quality of part suggestion results generated by cIMLE and cDDPM.}
\label{tab:quantitative-comparison}
\small
\centering
\begin{tabular}{ llcccc }
\toprule
Category & Method & VD & SD & VQ & PQ \\
\midrule
\multirow{2}*{PML chairs} & cIMLE & $\mathbf{3.11}$ & $\mathbf{2.91}$ & $2.82$ & $3.09$ \\
 & cDDPM & $3.07$ & $1.92$ & $\mathbf{3.70}$ & $\mathbf{3.84}$ \\
\midrule
\multirow{2}*{ShapeNet chairs} & cIMLE & $\mathbf{2.98}$ & $\mathbf{2.05}$ & $2.55$ & $2.83$ \\
 & cDDPM & $2.92$ & $1.73$ & $\mathbf{3.09}$ & $\mathbf{3.36}$ \\
\midrule
\multirow{2}*{ShapeNet airplanes} & cIMLE & $\mathbf{3.11}$ & $\mathbf{2.28}$ & $2.72$ & $3.22$ \\
 & cDDPM & $2.51$ & $1.76$ & $\mathbf{3.10}$ & $\mathbf{3.51}$ \\
\midrule
\multirow{2}*{ShapeNet guitars} & cIMLE & $\mathbf{3.42}$ & $\mathbf{1.92}$ & $2.77$ & $2.95$ \\
 & cDDPM & $2.21$ & $1.49$ & $\mathbf{3.18}$ & $\mathbf{3.65}$ \\
\bottomrule
\end{tabular}
\end{table}

Since cIMLE and cDDPM perform considerably better than the other two methods, we further compare both the diversity and quality of the suggested parts generated by cIMLE and cDDPM with a more detailed quantitative evaluation involving visual inspection of the results. We first choose $100$ initial parts from each category and use cIMLE and cDDPM to generate $4$ suggested parts for each initial part. Then, we visually inspect the generated parts and determine the following measures: visual diversity (VD), semantic diversity (SD), visual quality (VQ), and positioning quality (PQ). VD denotes the count of visually diverse part instances, SD stands for the number of semantic labels associated with the parts, VQ represents the number of parts that have high visual quality, and PQ indicates the count of parts that are positioned reasonably well in relation to the initial part. Since $4$ parts were generated, the counts can lie between $0$ and $4$. We present in Table~\ref{tab:quantitative-comparison} the averages of the aforementioned metrics for the suggested parts generated by each method across all datasets. We observe that, among the $4$ suggested parts, cIMLE provides parts with better diversity, while the parts generated by cDDPM possess higher quality.

\subsection{Shape synthesis}\label{res:synth}

Here, we present the experiments of synthesizing new shapes using our method. We also compare our synthesized shapes to the shapes generated by other 3D generative models.

\begin{figure*}[t!]
    \centering
    \begin{subfigure}{0.33\linewidth}
        \centering
        \includegraphics{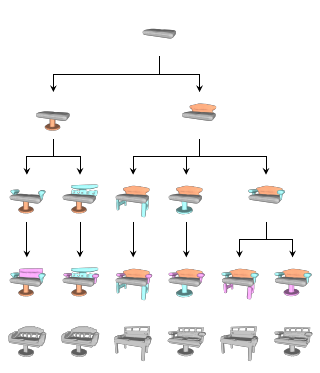}
        \caption{PML chairs}
        \label{fig:iterative-cimle-pml-chairs}
    \end{subfigure}
    \hfill
    \begin{subfigure}{0.275\linewidth}
        \centering
        \includegraphics{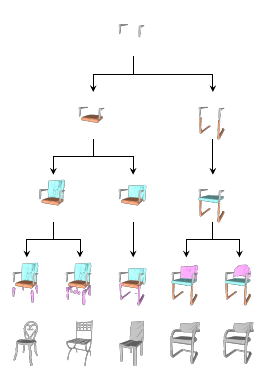}
        \caption{ShapeNet chairs}
        \label{fig:iterative-cimle-shapenet-chairs}
    \end{subfigure}
    \hfill
    \begin{subfigure}{0.385\linewidth}
        \centering
        \includegraphics{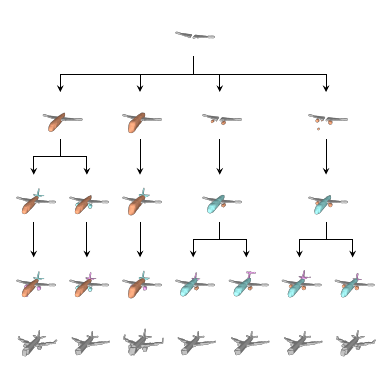}
        \caption{ShapeNet airplanes}
        \label{fig:iterative-cimle-shapenet-airplanes}
    \end{subfigure}
    \caption{Examples of shapes synthesized with cIMLE through part suggestion. Starting from an initial part (in gray at the top row), our system incrementally suggests possible parts to complement the current partial shape, until a predefined number of iterations (e.g., $4$) is reached. At each iteration, the user can select one of the suggested parts and connect it to the existing part(s). In this manner, various shapes can be synthesized. We can see that the synthesized shapes differ from their nearest neighbors in the training set (in gray at the bottom row).}
    \label{fig:iterative-cimle}
\end{figure*}

\begin{figure*}[t!]
    \centering
    \begin{subfigure}{0.275\linewidth}
        \centering
        \includegraphics{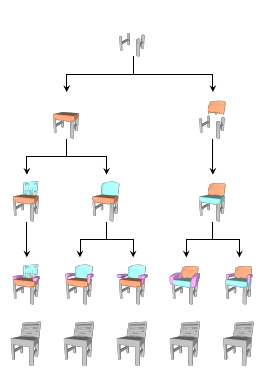}
        \caption{PML chairs}
        \label{fig:iterative-cddpm-pml-chairs}
    \end{subfigure}
    \hfill
    \begin{subfigure}{0.385\linewidth}
        \centering
        \includegraphics{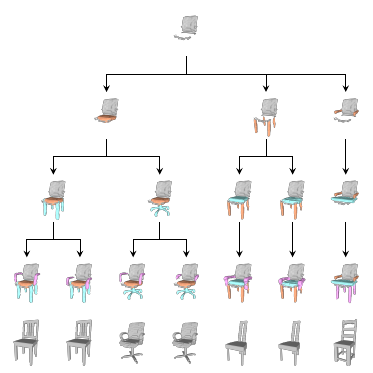}
        \caption{ShapeNet chairs}
        \label{fig:iterative-cddpm-shapenet-chairs}
    \end{subfigure}
    \hfill
    \begin{subfigure}{0.33\linewidth}
        \centering
        \includegraphics{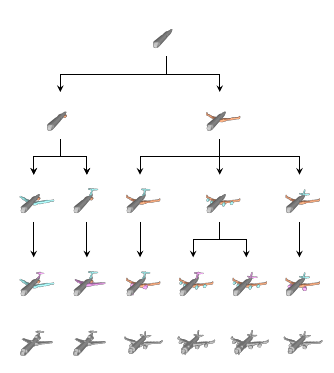}
        \caption{ShapeNet airplanes}
        \label{fig:iterative-cddpm-shapenet-airplanes}
    \end{subfigure}
    \caption{Examples of shapes synthesized with cDDPM through part suggestion. The setting of our incremental part suggestion system is the same as described in Figure~\ref{fig:iterative-cimle}. The nearest neighbors from the training set are presented in gray under each of the synthesized shapes.}
    \label{fig:iterative-cddpm}
\end{figure*}

\paragraph{Incremental part suggestion} Figure~\ref{fig:iterative-cimle} and Figure~\ref{fig:iterative-cddpm} showcase example shapes produced by our method in an incremental manner based on different training data (the PML chairs, ShapeNet chairs, and ShapeNet airplanes), employing the best-performing methods cIMLE and cDDPM as implementations of the PSN, respectively. We illustrate the creation of the synthesized shapes as a spanning tree, where the root node represents the initial part from which the synthesis process starts, each inner node simulates the user's selection of a part suggestion, and the leaf nodes represent the shapes synthesized on different spanning paths. Specifically, to synthesize these shapes, we first randomly sample a latent vector from the latent space learned by the PCN and then decode the latent vector into the initial part using the implicit decoder; see the initial parts in gray in Figure~\ref{fig:iterative-cimle} and Figure~\ref{fig:iterative-cddpm}. In an interactive setting, the user can handpick the initial part. Then, our system starts to provide part suggestions in an iterative manner. Given the current shape part(s), the PSN generates multiple latent samples that are reconstructed into suggested parts by the implicit decoder. The user manually selects suggested parts that are reasonable, and our system connects the selected part to the current part(s). The newly assembled parts are given to our system for generating part suggestions in the next iteration. In our experiments, the iterative synthesis process ends after a predefined number of iterations, although in an interactive setting the user can end the process earlier when satisfied with the synthesized shape.

We see from the results that our method using cIMLE or cDDPM is able to synthesize shapes with variations in both their overall structure and local geometry. Under each synthesized shape in Figure~\ref{fig:iterative-cimle} and Figure~\ref{fig:iterative-cddpm}, we present its nearest neighbor in the training set based on ED. We see that the synthesized shapes are different from the existing training shapes. For example, in Figure~\ref{fig:iterative-cimle-pml-chairs}, starting from a chair seat, cIMLE suggests diverse options for chair backs and legs. Meanwhile, in Figure~\ref{fig:iterative-cddpm-pml-chairs}, cDDPM generates various chair backs and arms given the initial chair legs. The synthesized shapes in Figure~\ref{fig:iterative-cimle-shapenet-chairs} showcase a variety of chair backs and those in Figure~\ref{fig:iterative-cddpm-shapenet-chairs} present different types of chair legs. Lastly, in Figure~\ref{fig:iterative-cimle-shapenet-airplanes}, we see variations in airplane bodies and tails, while in Figure~\ref{fig:iterative-cddpm-shapenet-airplanes}, we can observe different designs of airplane tails and engines.

\begin{figure*}
    \centering
    \begin{subfigure}{0.44\linewidth}
        \centering
        \includegraphics{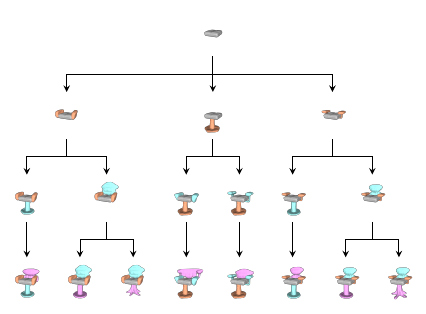}
        \caption{Initial width $\times0.5$}
        \label{fig:edit-width-0.5}
    \end{subfigure}
    \hfill
    \begin{subfigure}{0.55\linewidth}
        \centering
        \includegraphics{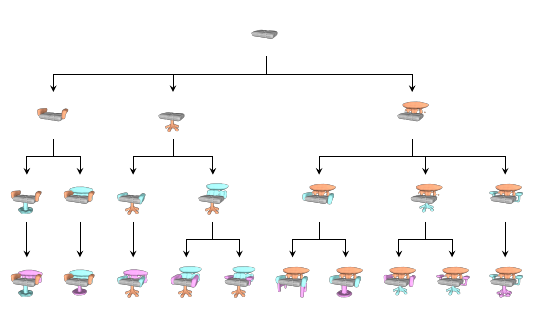}
        \caption{Initial width $\times0.75$}
        \label{fig:edit-width-0.75}
    \end{subfigure}
    \caption{Examples of shapes synthesized from edited initial parts, where the original initial parts are shown in Figure~\ref{fig:iterative-cimle-pml-chairs}.}
    \label{fig:edit}
\end{figure*}

Our iterative part suggestion also allows synthesizing shapes with edited initial parts. Figure~\ref{fig:edit} shows shapes synthesized starting from edited versions of the chair seats given in Figure~\ref{fig:iterative-cimle-pml-chairs}, where we multiply the width of the initial part by $0.5$ in Figure~\ref{fig:edit-width-0.5} and by $0.75$ in Figure~\ref{fig:edit-width-0.75}. We see that our method can generate parts that adapt to the user's modification of the initial parts, e.g., the suggested chair backs and arms have similar widths as the initial parts and are placed at reasonable positions. Note that we base this experiment on the synthetic shapes generated by the PML, as the synthetic training data with gradual changes in shape attributes enables the learning of a smoother and denser latent manifold~\cite{ali2021evaluation} which allows the PSN to generate more meaningful part suggestions.

\paragraph{Comparison to 3D generative baselines} We qualitatively and quantitatively compare our synthesized results to the shapes generated by two state-of-the-art 3D generative models, namely, IM-NET~\cite{chen2019learning} and PQ-NET~\cite{wu2020pq}. We compare to these two methods as they are representatives of:
\begin{enumerate*}[label=(\roman*)]
\item a method that synthesizes shapes as a whole and
\item a method that synthesizes shapes in a part-based representation.
\end{enumerate*}
We use their released code and train their models using our data. All the results are sampled under the resolution of $64\times64\times64$. Both IM-NET and PQ-NET perform shape generation by randomly sampling from the latent space learned by a latent GAN. The latent GAN of IM-NET is trained with the latent vectors encoded from the entire shapes, while the latent space of PQ-NET is based on a part AE. More similarly to our method, PQ-NET further performs part generation and assembly based on recurrent neural networks.

\begin{figure*}[t!]
    \centering
    \begin{subfigure}{0.495\linewidth}
        \centering
        \includegraphics{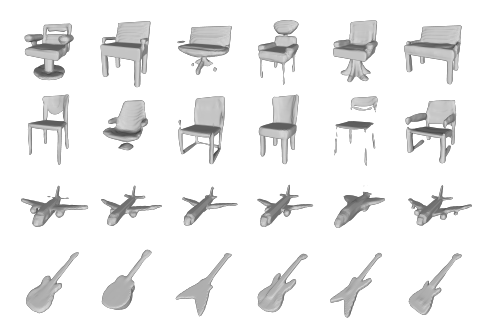}
        \caption{IM-NET}
        \label{fig:synthesis-imnet}
    \end{subfigure}
    \hfill
    \begin{subfigure}{0.495\linewidth}
        \centering
        \includegraphics{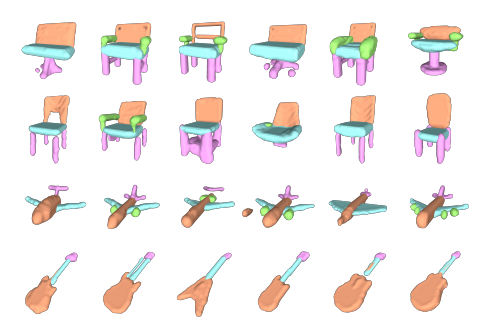}
        \caption{PQ-NET}
        \label{fig:synthesis-pqnet}
    \end{subfigure}
    \\[1.5ex]
    \begin{subfigure}{0.495\linewidth}
        \centering
        \includegraphics{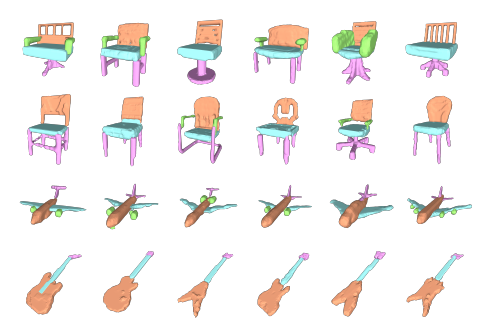}
        \caption{Ours (cIMLE)}
        \label{fig:synthesis-cimle}
    \end{subfigure}
    \hfill
    \begin{subfigure}{0.495\linewidth}
        \centering
        \includegraphics{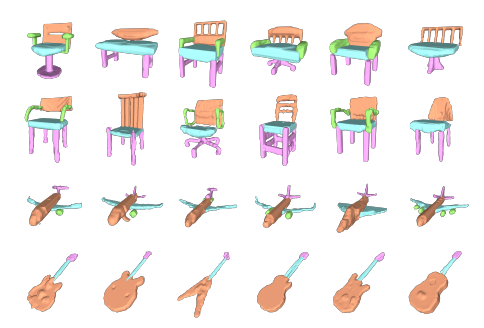}
        \caption{Ours (cDDPM)}
        \label{fig:synthesis-cddpm}
    \end{subfigure}
    \caption{Our shape synthesis results using cIMLE and cDDPM as the implementation of the PSN, compared to the shapes generated by IM-NET and PQ-NET.}
    \label{fig:synthesis}
\end{figure*}

Figure~\ref{fig:synthesis} visually compares our results to those of IM-NET and PQ-NET, where we manually selected reasonable samples for each method. In each subfigure, we present from top to bottom shapes synthesized based on the PML chairs, ShapeNet chairs, ShapeNet airplanes, and ShapeNet guitars. Our user-in-the-loop shape synthesis method possesses several advantages over the GAN-based generative methods. For example, the user can rule out unreasonable shape parts during their selection, and thus the shapes synthesized by our method have more visually plausible parts, while it is difficult to ensure a perfect shape with the GANs given that shapes are generated as a whole. For example, randomly sampling with the GANs can produce results with disconnected chair legs (Figure~\ref{fig:synthesis-imnet}) or floating blobs on the side of the airplane bodies or around chair legs (Figure~\ref{fig:synthesis-pqnet}). Moreover, with the better reconstruction quality, our synthesized shapes preserve more geometric details in the generated parts.

\begin{table}[t!]
\caption{Quantitative evaluation of shape synthesis results via different methods, where MMD is multiplied by $10^3$ and JSD is multiplied by $10^2$.}
\label{tab:shape-synthesis}
\small
\centering
\begin{tabular}{ llccc }
\toprule
Category & Method & COV & MMD & JSD \\
\midrule
\multirow{4}*{PML chairs} & IM-NET & $\SI{52}{\percent}$ & $9.241$ & $5.299$ \\
 & PQ-NET & $\SI{49}{\percent}$ & $7.865$ & $3.816$ \\
 & Ours (cIMLE) & $\mathbf{\SI{55}{\percent}}$ & $\mathbf{7.654}$ & $\mathbf{2.869}$ \\
 & Ours (cDDPM) & $\SI{53}{\percent}$ & $8.032$ & $3.081$ \\
\midrule
\multirow{4}*{ShapeNet chairs} & IM-NET & $\SI{41}{\percent}$ & $11.76$ & $5.917$ \\
 & PQ-NET & $\SI{40}{\percent}$ & $10.37$ & $5.396$ \\
 & Ours (cIMLE) & $\SI{47}{\percent}$ & $\mathbf{10.19}$ & $3.898$ \\
 & Ours (cDDPM) & $\mathbf{\SI{48}{\percent}}$ & $11.08$ & $\mathbf{3.699}$ \\
\midrule
\multirow{4}*{ShapeNet airplanes} & IM-NET & $\SI{48}{\percent}$ & $4.909$ & $6.147$ \\
 & PQ-NET & $\SI{45}{\percent}$ & $4.889$ & $6.851$ \\
 & Ours (cIMLE) & $\SI{52}{\percent}$ & $5.054$ & $6.098$ \\
 & Ours (cDDPM) & $\mathbf{\SI{54}{\percent}}$ & $\mathbf{4.713}$ & $\mathbf{5.722}$ \\
\midrule
\multirow{4}*{ShapeNet guitars} & IM-NET & $\SI{41}{\percent}$ & $2.213$ & $3.178$ \\
 & PQ-NET & $\SI{42}{\percent}$ & $2.022$ & $3.224$ \\
 & Ours (cIMLE) & $\mathbf{\SI{53}{\percent}}$ & $\mathbf{1.756}$ & $\mathbf{2.583}$ \\
 & Ours (cDDPM) & $\SI{49}{\percent}$ & $1.829$ & $2.703$ \\
\bottomrule
\end{tabular}
\end{table}

To quantitatively evaluate the results, we adopt the coverage (COV), minimum matching distance (MMD), and Jensen--Shannon divergence (JSD) as evaluation metrics~\cite{achlioptas2018learning}, where COV and JSD evaluate the diversity of the results, and MMD evaluates the fidelity. In the calculation of COV and MMD, we use CD as the difference metric, and to enhance the distinction of MMD values, we base the computation of CD on the squared ED computed between each pair of points. For each method, we collect a set of shapes having $100$ samples and compare to the test set using the aforementioned metrics. For each shape, we reconstruct a corresponding triangle mesh and sample $2048$ points on the mesh surface to compute CD. Table~\ref{tab:shape-synthesis} details this quantitative evaluation. Our method consistently outperforms others across all metrics and shape categories. The performance of our method varies when employing different models for implementing the PSN. When using cIMLE as the implementation of the PSN, our method achieves the highest diversity and fidelity in synthesizing shapes from the PML chairs and ShapeNet guitars, as well as the highest fidelity on the ShapeNet chairs. When using cDDPM, our method attains the highest diversity and fidelity on the ShapeNet airplanes and the highest diversity on the ShapeNet chairs. We note that PQ-NET, which also operates at the part level for shape synthesis, performs worse than IM-NET in COV and JSD for some shape categories. We hypothesize that this is due to our relatively small training set, lowering PQ-NET's ability to generate diverse parts and potentially leading to the repetition of certain parts in different shapes.

\begin{figure}[t!]
    \centering
    \begin{subfigure}{0.33\linewidth}
        \centering
        \includegraphics{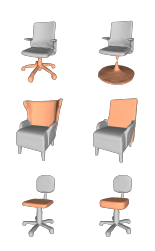}
        \caption{ANISE}
        \label{fig:completion-anise}
    \end{subfigure}
    \hfill
    \begin{subfigure}{0.66\linewidth}
        \centering
        \includegraphics{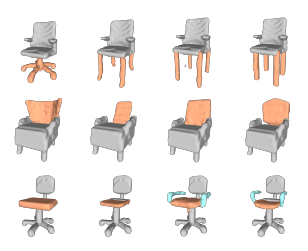}
        \caption{Ours}
        \label{fig:completion-ours}
    \end{subfigure}
    \caption{Shape completion and editing obtained with ANISE and our method, based on the provided partial assembly in gray.}
    \label{fig:completion}
\end{figure}

\paragraph{Comparison to ANISE~\cite{petrov2023anise}} Additionally, we conduct a qualitative comparison between our method and a recent method proposed by Petrov et al.~\cite{petrov2023anise}, named ANISE, that focuses on reconstructing shapes as implicit fields in a part-based fashion, while also allowing a limited form of shape synthesis through shape completion and part editing. We present this comparison in Figure~\ref{fig:completion}, where ANISE enables shape completion by producing a particular part to complement the given partial assembly, reconstructing a specific query shape (the first column in Figure~\ref{fig:completion-anise}). It then allows users to interpolate the part to edit the reconstruction (the second column in Figure~\ref{fig:completion-anise}). On the other hand, our method is able to directly provide diverse part suggestions according to the partial shape, without the need of a reference shape (Figure~\ref{fig:completion-ours}). In this comparison, the implementation of the PSN is based on cDDPM, which enables the sampling of a variable number of part latent codes and thus maximizes the diversity of the generated parts. 

We observe from Figure~\ref{fig:completion} that, compared to ANISE, our method provides a variety of parts to complement the same partial shape. Moreover, due to the nature of incremental shape synthesis, our method has the potential to extend the completed shape by generating new parts on top of the previously generated ones; see the chair arms in the last two examples of Figure~\ref{fig:completion-ours}. Note that, since our method prioritizes the diversity of the synthesized shapes over reconstruction details, we only sample our results under the resolution of $64\times64\times64$. Consequently, our results exhibit blurrier geometric details and feature less smoothed surfaces compared to those generated by ANISE.

\section{Conclusion, limitations, and future work}\label{sec:concl}

We introduced a method that iteratively synthesizes a shape by suggesting parts that can complement a partial assembly. Our method synthesizes the geometry of the parts in an implicit representation and positions the parts relative to the partial assembly. Using this method, we conducted a comparative study of different approaches for multimodal generation and showed with different forms of evaluation that our method, implemented with the best-performing multimodal generative models, outperforms existing deep learning-based shape synthesis approaches in terms of synthesis diversity and quality.

\begin{wrapfigure}{r}{0.38\linewidth}
    \centering
    \includegraphics{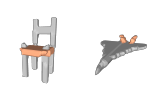}
    \caption{Misplaced parts (in orange) predicted by the PCN.}
    \label{fig:limitation}
\end{wrapfigure}

Our method has a few limitations arising from our network design. The PCN predicts the position of synthesized parts based on their geometry. Thus, in some cases, e.g., if a part is significantly different from the training samples, the PCN may predict a less than ideal placement. Figure~\ref{fig:limitation} shows a few examples, where the parts with a failure in their placement are in orange. We can see that the chair seat and the airplane tail are placed at slightly inaccurate positions by the PCN. To quantify the occurrence of misaligned parts, we use the PCN to reconstruct the test set of each category of shapes, each test set comprising $100$ shapes not encountered during training. Following a careful visual inspection, we identify $29$ instances of the ShapeNet chairs and $27$ instances of the ShapeNet airplanes having notably misplaced parts in their reconstructions, while the reconstructions of the PML chairs and ShapeNet guitars do not suffer from this issue. In addition, the quality of the part suggestions depends on the diversity present in the training data. For example, certain types of shapes in the ShapeNet dataset do not posses a great variety of parts, and thus part suggestions for these types of shapes are also limited. In contrast, we obtain a richer variety of part suggestions for shapes in the PML dataset, since the training samples have more variety and gradual differences.

In terms of future work, it may be interesting to replace our part-level AE with an AE that predicts a latent vector for an entire input shape and then predict latent vectors of individual parts from the per-shape latent vector~\cite{dubrovina2019composite,hertz2022spaghetti}. This would make reconstruction more straightforward and combine the advantages of per-shape and per-part representations into a single method. Moreover, due to the design of the PCN, our method is currently limited to shapes decomposed into semantic parts. It would be interesting to adapt our method to datasets with a finer shape decomposition, such as PartNet~\cite{mo2019partnet}.

\section*{Acknowledgments}

We thank the anonymous reviewers for their valuable comments and suggestions. This work was supported in part by the Natural Sciences and Engineering Research Council of Canada through a Discovery Grant (2022-04903).

\bibliographystyle{elsarticle-num}
\bibliography{references}

\end{document}